\documentclass[12pt,a4paper]{article}

\usepackage{latexsym}
\usepackage{amsmath}
\usepackage{amsfonts}
\usepackage{amssymb}

\newcommand{\be}{\begin{equation}}
\newcommand{\ee}{\end{equation}}
\newcommand{\bea}{\begin{eqnarray}}
\newcommand{\eea}{\end{eqnarray}}

\def\cO{{\cal O}}                       %
\def\ri{{\mathrm{i}}}                   %
\def\bR{{\mathbb R}}                    %
\def\bC{{\mathbb C}}                    %
\def\1{{\mbox{\boldmath $1$}}}          %

\def\0{{\mbox{\boldmath $0$}}}          %
\def\cF{{\cal F}}                       %
\def\cD{{\cal D}}                       %
\def\fH{{\mathfrak H}}                       %

\def\cI{{\mathcal I}}                    %
\def\cC{{\mathcal C}}                    %
\def\cM{{\mathcal M}}                    %
\def\cD{{\mathcal D}}                    %
\def\cH{{\mathcal H}}                    %
\def\bR{{\mathbb R}}                    %
\def\bZ{{\mathbb Z}}                    %
\def\ri{{\mathrm{i}}}                   %
\def\tr{\mathrm{tr}}                  %
\def\diag{\mathrm{diag}}                %
\def\cF{{\mathcal F}}                    %

\begin{document}

\vspace*{0.5cm}
\begin{center}
{\Large \bf On the superintegrability of the rational
Ruijsenaars-Schneider model}
\end{center}

\vspace{0.2cm}

\begin{center}
V.~Ayadi${}^{a}$
and L. Feh\'er${}^{a,b}$ \\

\bigskip

${}^a$Department of Theoretical Physics, University of Szeged\\
Tisza Lajos krt 84-86, H-6720 Szeged, Hungary\\
e-mail: ayadi.viktor@stud.u-szeged.hu

\bigskip

${}^{b}$Department of Theoretical Physics, MTA  KFKI RMKI\\
H-1525 Budapest, P.O.B. 49,  Hungary \\
e-mail: lfeher@rmki.kfki.hu

\bigskip

\end{center}

\vspace{0.2cm}

\begin{abstract}
The  rational and  hyperbolic
Ruijsenaars-Schneider models  and
their non-relativistic limits
are maximally superintegrable
since they admit action variables with globally
well-defined canonical conjugates.
In the case of the rational
Ruijsenaars-Schneider model we present an alternative proof
of the superintegrability  by explicitly exhibiting
extra conserved quantities
relying on a generalization of the construction of Wojciechowski
for the rational Calogero model.



\end{abstract}

\newpage

\section{Introduction}
\setcounter{equation}{0}

Let us consider a Hamiltonian system $(M, \omega, h)$, where $(M,
\omega)$ is a symplectic manifold of dimension $2n$ and $h$ is the
Hamiltonian. The system is \emph{Liouville integrable} if there
exist $n$  independent functions $h_i \in C^\infty(M)$ ($i=1,\ldots,
n)$ that are in involution with respect to the Poisson bracket and
the Hamiltonian $h$ is equal to one of the $h_i$.
A Liouville integrable system is called
\emph{maximally superintegrable}  if it admits $(n-1)$ additional
constants of motion, say $f_j\in C^\infty(M)$, that are
time-independent, globally smooth, and the $(2n-1)$  functions
\be
 h_1, \ldots, h_n, f_1, \ldots, f_{n-1}
\label{1}\ee
are independent, i.e.,
their differentials are linearly independent on a dense  submanifold of $M$.
As a general reference on superintegrability, we refer to \cite{CRM}.
Maximal superintegrability and the compactness of the joint level surfaces
\be
h_i= h_i^0 \quad (\forall i=1,\ldots,  n) \quad \hbox{and} \quad f_j=f_j^0
\quad
(\forall j=1,\ldots, n-1)
\label{2}\ee
for generic constants $h_i^0$ and $f_j^0$
implies the periodicity of the generic flows of the system $(M,\omega, h)$.
Maximally superintegrable systems with periodic flows are very rare,
the classical examples being the isotropic harmonic oscillator  and
the negative energy sector of the Kepler problem.
Other examples are provided by magnetic analogues and higher dimensional generalizations
of the Kepler problem \cite{Rag}.
See also \cite{Evans, Mill} for interesting classification results in low dimensions.

There exists a large class of Liouville integrable systems that are
maximally superintegrable in a rather obvious manner.
These are the systems that admit action variables with
globally well-defined canonical conjugates.
Typical examples are  scattering systems having configuration space trajectories
$q(t) = (q_1(t),\ldots, q_n(t))$ with large time asymptotes of the form
\be
q_i(t) \sim q_i^+ +  t v_i^+
\quad (\forall i=1,\ldots, n)
\quad
\quad\hbox{as}\quad t \to \infty
\label{3}\ee
in such a way that the action variables $p_i^+=m_i v_i^+$ and their canonical conjugates
belong to $C^\infty(M)$ and together they parametrize the phase space $M$.
This is the expected behaviour  in
many-body models of particles interacting via repulsive pair potentials.
For example,
the scattering characteristics of the rational and hyperbolic Calogero-Sutherland
models \cite{Cal,Sut,CalRag} as well as of their Ruijsenaars-Schneider (RS) deformations
\cite{RS}
have been analyzed in \cite{SR-CMP}, and the results
imply  that these  many-body models are maximally
superintegrable.

In our opinion, it is interesting to find explicit expressions for
the constants of motion even for those systems whose maximal
superintegrability is already known from abstract arguments.
This is especially true if simple expressions can be obtained.
A nice example is the  direct proof of the
maximal superintegrability of the rational Calogero model
presented by Wojciechowski \cite{W}.
His arguments were later generalized  to the rational Calogero models based
on arbitrary finite Coxeter groups \cite{Sas}.
An attempt to study the hyperbolic Sutherland model  in a similar manner
was made in \cite{Gon}, but in this case the constants of motion  have a more complicated structure
and to our knowledge  fully explicit globally smooth expressions of them are still not available.

The main goal of this letter is to  explicitly exhibit constants of motion that
show the maximal superintegrability of the rational Ruijsenaars-Schneider  model.
We find that  the method of Wojciechowski can be successfully applied in this case.
Before explaining this in Section 3,  in the next section we  present a general argument
implying the maximal superintegrability of the Liouville integrable systems that admit
action variables with globally well-defined canonical conjugates.
The precise definition of the foregoing condition is given below.

\section{A class of maximally superintegrable systems}

Consider the Liouville integrable systems
$(M, \omega, h_i)$ associated with the Poisson commuting,  independent
Hamiltonians $h_1, \ldots, h_n$.
Let us assume that globally well-defined action variables
with globally well-defined canonical conjugates exist.
By definition,
this means that we have a phase space $(\cM, \Omega)$
of the form
\be
\cM:= \bR^n \times \cD_n = \{ (Q,P)\,\vert\, Q\in \bR^n,\, P\in \cD_n\}
\label{4}\ee
with a connected open domain $\cD_n\subseteq \bR^n$ and
canonical symplectic form
\be
\Omega = \sum_{i=1}^n dP_i \wedge d Q_i,
\label{5}\ee
which is symplectomorphic to $(M, \omega)$ and permits
identification of the Hamiltonians $\{h_i\}$ as functions of the action variables $\{P_j\}$.
More precisely, we assume the existence of a symplectomorphism
\be
A: M\to \cM
\label{6}\ee
such that the functions $\cH_i := h_i \circ A^{-1}$ do not depend on the variables $\{Q_j\}$ and
\be
X_{i,j}:= \frac{\partial \cH_i}{\partial P_j}
\label{7}\ee
yields an invertible matrix $X(P)$ at every  $P\in \cD_n$.
The map $A$ is referred to as a global action-angle map of maximally
non-compact type.

If a global action-angle map of the above type exists,
then one can introduce the functions $f_i \in C^\infty(M)$ ($i=1,\ldots,n)$ by the definition
\be
(f_i \circ A^{-1})(Q,P) := \sum_{j=1}^n Q_j X(P)^{-1}_{j,i}
\quad
\hbox{with}\quad
\sum_{j=1}^n X(P)_{i,j} X(P)^{-1}_{j,k} = \delta_{i,k}.
\label{8}\ee
By using that $A$ is a symplectomorphism, one  obtains the Poisson brackets
\be
\{ f_i, h_j\}_{M} = \delta_{i,j},
\qquad
\{ f_i, f_j\}_{M}=0.
\label{9}\ee
Together with $\{ h_i, h_j\}_{M}=0$, these imply that
the $2n$ functions $h_1,\ldots, h_n, f_1, \ldots, f_n$ are functionally independent
at every point of $M$.
The choice of any of these $2n$ functions as the Hamiltonian gives rise to a maximally
superintegrable system.
For example, the $(2n-1)$ independent functions  $h_1,\ldots, h_n, f_2, \ldots, f_n$
(resp.~$h_2, \ldots, h_n, f_1,\ldots, f_n$) Poisson commute with $h_1$
(resp.~with $f_1$).

Let us now study a Hamiltonian $H\in C^\infty(M)$ of the form
$H=\fH(h_1,\ldots, h_n)$ with some $\fH \in C^\infty(\bR^n)$.
Under mild conditions, this Hamiltonian is also maximally superintegrable.
For example, suppose that there exists an index
$l\in \{1,\ldots, n\}$ such that $\partial_l \fH  \neq 0$ generically.
For $l=1$, the assumption guarantees  that $H, h_2,\ldots, h_n$ are independent,
i.e., $H$ is Liouville integrable in the sense specified at the beginning.
Next, let us choose  $(n-1)$ smooth maps $V_a: \bR^n \to \bR^n$ ($a=1,\ldots, n-1$) that satisfy
the equations
\be
\sum_{k=1}^n V_a^k \partial_k \fH =0
\label{E1}\ee
identically on $\bR^n$, and their values yield independent $\bR^n$-vectors
generically.
In other words,  the vectors $V_a(x)$ span the orthogonal complement of the gradient of
$\fH$ at generic points $x\in \bR^n$.
Such maps $V_a$ always exist, since the independence
of their values is required only generically.
By using $V_a$, we define the function $F_a\in C^\infty(M)$  by
\be
F_a:= \sum_{k=1}^n f_k V_a^k(h_1,\ldots, h_n), \qquad
\forall a=1,\ldots, n-1.
\label{E2}\ee
It is easily seen that the globally smooth functions $H, h_2, \ldots, h_n, F_1, \ldots, F_{n-1}$
are independent  and they Poisson commute with $H$.
This demonstrates that $H$ is maximally superintegrable.
Incidentally, the set $h_1,\ldots, h_n, F_1, \dots, F_{n-1}$ also gives $(2n-1)$
independent constants of motion for $H$, and this holds  even if we drop
our technical assumption on the gradient of $\fH\in C^\infty(\bR^n)$.

The above arguments are quite obvious and are well known to experts (see e.g.~\cite{Tsig}).
Among their consequences, we wish to stress the fact that
\emph{all Calogero type models possessing only scattering trajectories
are maximally superintegrable}.
Indeed, for these models action-angle maps of
maximally non-compact type were constructed by Ruijsenaars \cite{SR-CMP}.
Concretely, this is valid for the rational Calogero model,
the hyperbolic Sutherland model, and for the relativistic deformations of these
models due to Ruijsenaars and Schneider.
The construction of the pertinent action-angle maps relies on algebraic
procedures, but  fully explicit
formulae are not available.
Therefore it might be interesting to display the maximal superintegrability of
these models by alternative direct constructions of the required constants of motion.

\section{Constants of motion in the rational RS model}

Let us recall that the phase space of the rational Ruijsenaars-Schneider model is
\be
M =
 \cC_n \times \bR^n = \{ (q,p)\,\vert\, q\in \cC_n,\, p\in \bR^n\}
\label{10}\ee
where
\be
\cC_n := \{ q\in \bR^n\,\vert\, q_1 > q_2 > \cdots >q_n\}.
\label{11}\ee
The symplectic structure $\omega= \sum_{k=1}^n dp_k \wedge d q_k$
corresponds to the fundamental Poisson brackets
\be
\{ q_i, p_j\}_{M}= \delta_{i,j},
\qquad
\{ q_i, q_j\}_M = \{ p_i, p_j\}_M =0.
\label{12}\ee
The commuting Hamiltonians of the model
are generated by the (Hermitian, positive definite)  Lax matrix \cite{RS}  given by
\be
L(q,p)_{j,k}=u_j(q,p)
 \left[ \frac{\ri \chi}{\ri \chi + (q_j-q_k)} \right] u_k(q,p)
\label{13}\ee
with the $\bR_+$-valued functions
\be
u_j(q, p):=  e^{p_j} \prod_{m\neq j}
 \left[ 1 + \frac{ \chi^2}{(q_j - q_m)^2}\right]^\frac{1}{4},
 \label{14}\ee
 where $\chi\neq 0$ is an arbitrary real coupling constant
 and the `velocity of light' is set to unity.

 By using the diagonal matrix $\mathbf{q} :=\diag(q_1,\ldots, q_n)$,
we define the functions $I_k, I_k^1 \in C^\infty(M)$ by
 \be
 I_k(q,p):= \tr\left( L(q,p)^k\right),
 \quad
 I_k^1(q,p):= \tr\left(\mathbf{q}  L(q,p)^k \right)
 \qquad \forall k\in\bZ.
 \label{15}\ee
 It is well-known that the functions $I_k$ pairwise Poisson commute
 and a convenient generating set of the spectral invariants of $L$
 is provided by the independent functions $I_1, \ldots, I_n$.
 Note that one may also use the coefficients of the characteristic
 polynomial of $L(q,p)$ as an alternative generating set.
 The `principal Hamiltonian' of the model is
 \be
 h = \frac{1}{2}( I_1 + I_{-1}) = \sum_{k=1}^n \cosh(p_k) \prod_{j\neq k}\left[ 1+
\frac{\chi^2}{(q_k - q_j)^2}\right]^\frac{1}{2}.
 \label{17}\ee
 Our subsequent considerations are based on the following important formula:
 \be
 \{ I_k^1, I_j\}_M = j I_{j+k}
 \qquad
 \forall j,k\in \bZ.
 \label{18}\ee
 This generalizes an analogous formula found by Wojciechowski \cite{W}
 in the rational Calogero model.
The quantities $I_j$ and $I_k^1$ form an infinite dimensional Lie algebra under the Poisson bracket,
with the $I_k^1$ realizing the centerless Virasoro algebra:
\be
\{ I_k^1, I_j^1\}_M= (j-k) I_{k+j}^1
\qquad
\forall j,k\in \bZ.
\label{19}\ee
We postpone the proof of the above relations for a little while.

Since $\dim(M)=2n$ is finite, only finitely many of the functions $I_k$, $I_m^1$ can be independent.
A set of $2n$ independent functions  is given, for example, by
\be
I_1,\ldots, I_n, I_1^1,\ldots, I_n^1.
\label{20}\ee
To see that the Jacobian determinant
\be
J:= \det \frac{\partial (I_1, \ldots, I_n, I_1^1,\ldots, I_n^1)}
{\partial (p_1,\ldots, p_n, q_1,\ldots, q_n)}
\label{21}\ee
is non-vanishing generically, notice that $J$ is the ratio of two polynomials in the
$2n$-variables $e^{p_i}, q_i$ ($i=1,\ldots, n$), and hence it either vanishes identically
or is non-vanishing on a dense submanifold of the phase space. It is easy to confirm the
non-vanishing of $J$ in the asymptotic region where the coordinate-differences are large.
In that region $L(q,p)$ (\ref{13}) becomes diagonal, which implies  the leading behaviour
\be
I_k \sim \sum_{i=1}^n e^{2 k p_i},
\quad
I_k^1 \sim \sum_{i=1}^n q_i e^{2 k p_i},
\label{22}\ee
whereby the leading term of  $J$ can be calculated in terms of Vandermonde determinants.
We conclude from the inverse function theorem that the $2n$ functions (\ref{20}) can be serve as
independent coordinates locally, around generic points of the phase space.

To elaborate the consequences of (\ref{18}),
let us first consider an arbitrary smooth function $I=I(I_1,\ldots, I_n)$.
Then we obtain from (\ref{18})
that $\{ \{ I_k^1, I\}_M, I\}_M =0$.
This entails that $I_k^1$ develops linearly along the Hamiltonian flow of $I$,
\be
I_k^1(q(t), p(t))= I_k^1(q(0), p(0)) + t \{ I_k^1, I\}_M(q(0), p(0)),
\qquad \forall k\in \bZ.
\label{23}\ee
In particular, the independent functions given in (\ref{20}) provide an algebraic linearization
of the flow of $I$ in the sense of \cite{Sas}, i.e., they can be used as
explicitly given alternatives
to  action-angle type variables.

Next, we observe that  the functions
\be
I_j^1\{I_k^1, I \}_M - I_k^1 \{I_j^1, I\}_M,
\qquad
\forall j,k\in \bZ,
\label{24}\ee
Poisson commute with $I$.
By using these,
it is possible to
exhibit  $(2n-1)$ smooth functions on $M$
that Poisson commute with $I=I(I_1,\ldots, I_n)$ and are functionally independent generically.
For instance, for any fixed $j\in \{1,2,\ldots, n\}$ consider the functions
\be
C_{k,j}:= I_k^1 I_{2j} - I_j^1 I_{k+j},
\qquad k\in \{ 1,2,\ldots, n\} \setminus \{j\}.
\label{25}\ee
By taking the $2n$ functions (\ref{20}) as coordinates around generic points of $M$,
we can easily compute the Jacobian determinant
\be
J_j:= \det \frac{\partial (I_a, C_{b,j})}
{\partial ( I_\alpha, I^1_\beta)}
\quad\hbox{with}\quad
a,\alpha \in \{1,\ldots, n\},\,\, b,\beta \in \{1,\ldots, n\}\setminus \{j\},
\label{26}\ee
and obtain that $J_j= (I_{2j})^{n-1}$, which is generically non-zero.
The fact that  the $(2n-1)$ functions furnished by $I_m$ ($m=1,\ldots, n$) and $C_{k,j}$ (\ref{25}) are
independent and Poisson commute with $I_j$ shows directly that $I_j$ is
maximally superintegrable.

As another example, take the standard Ruijsenaars-Schneider Hamiltonian $h$ in (\ref{17}).
The maximal superintegrability of $h$ is ensured by the `extra constants of motion'
\be
K_j:= I_j^1 ( I_2 - n) - I_{1}^1 (I_{j+1} - I_{j-1}), \qquad
 j=2,\ldots, n.
\ee
Indeed, one has $\{K_j, h\}_M=0$ and
\be
\det \frac{\partial (I_a, K_b)}
{\partial ( I_\alpha, I^1_\beta)}= (I_2 - n)^{n-1},
\qquad
a,\alpha \in \{1,\ldots, n\},\, b,\beta \in \{2,\ldots, n\},
\label{27}\ee
which guarantees that the $(2n-1)$ functions $I_a, K_b$ are independent.
Similarly, the momentum ${\mathcal{P}}:= \frac{1}{2}(I_1 - I_{-1})$ admits the functionally independent
extra constants of motion  given by
$L_j:= I_j^1 ( I_2 + n) - I_{1}^1 (I_{j+1} + I_{j-1})$
for $j=2,\ldots, n$.

For a general $I=I(I_1,\ldots, I_n)$, let us
consider functions $\cF_a\in C^\infty(M)$ of the form
\be
\cF_a := \sum_{k=1}^n I_k^1 U_a^k(I_1,\ldots, I_n),
\ee
with some smooth maps $U_a: \bR^n \to \bR^n$
($a=1,\ldots, n-1$) subject to the identity
\be
\sum_{k=1}^n\Bigl( \sum_{j=1}^n j I_{j+k} \frac{\partial I}{\partial I_j}\Bigr)  U_a^k  =0,
\ee
which guarantees that $\{ \cF_a, I\}_M =0$ (cf.~equations (\ref{E1}) and (\ref{E2})).
One can always choose the maps $U_a$ ($a=1,\ldots, n-1$)  in such a way that their values
yield  linearly independent $\bR^n$-vectors
at generic arguments.
Then the $(2n-1)$ functions
$I_1,\ldots, I_n, \cF_1,\ldots, \cF_{n-1}$ are independent, globally smooth and
Poisson commute with $I$.

To summarize, we have seen that the relation (\ref{18}) leads to an
explicit linearization of the Hamiltonian flow associated with any
$I=I(I_1,\ldots, I_n)$ and allows us to display the maximal superintegrability of
the Ruijsenaars-Schneider Hamiltonian $h$ (\ref{17})
in an explicit manner (and similarly for $\mathcal{P}$ and  $I_j$ for $j=1,\ldots, n$).
The above arguments also imply, among others,
the maximal superintegrability of any polynomial Hamiltonian $I=I(I_1,\ldots, I_n)$.

\medskip

Now we prove the relations (\ref{18}) and (\ref{19}).
Direct verification is possible in principle, but it would require non-trivial
calculations.
 However, it is quite easy to obtain the claimed relations  by utilizing the
 derivation of the rational RS model in the symplectic reduction
 framework presented in \cite{JPA}.  To explain this,
we start by recalling from \cite{JPA}  the relevant
reduction of the phase space
\be
T^* GL(n,\bC) \times \cO(\chi) \equiv GL(n,\bC) \times gl(n,\bC) \times \cO(\chi) = \{(g, J^R, \xi)\}.
\label{A1}\ee
Here $\cO(\chi)$ is a minimal coadjoint orbit of the group $U(n)$, which as a set is given by
\be
\cO({\chi}):= \{ \ri\chi (\1_n- v v^\dagger)   \,\vert\, v\in \bC^n,\, \vert v \vert^2 = n\}.
\label{A2}\ee
In (\ref{A1}) we use the trivialization of $T^*GL(n,\bC)$ by left-translations and identify the
real Lie algebra $gl(n,\bC)$ with its dual space with the aid of the `scalar product'
provided by the real part of the trace
\be
\langle X, Y \rangle := \Re \tr(XY),
\qquad
\forall X,Y \in gl(n,\bC).
\label{A3}\ee
In terms of evaluation functions, the not identically zero fundamental
Poisson brackets read
\be
\{g, \langle X, J^R\rangle \} = g X,
\quad
\{ \langle X, J^R\rangle, \langle Y, J^R \rangle \} = -\langle [X,Y], J^R \rangle,
\quad
\{ \langle X_+, \xi \rangle, \langle Y_+, \xi \rangle \} = \langle [X_+,Y_+], \xi \rangle,
\label{A4}\ee
where $X_+$, $Y_+$ are the anti-hermitian parts of the constants
$X,Y\in gl(n,\bC)$.
The reduction is based on using the symmetry group
$K:= U(n) \times U(n)$, where an element $(\eta_L, \eta_R) \in K$ acts via
the symplectomorphism $\Psi_{(\eta_L, \eta_R)}$ defined by
\be
\Psi_{(\eta_L,\eta_R)}: (g,J^R,\xi) \mapsto (\eta_L g \eta_R^{-1}, \eta_R J^R \eta_R^{-1}, \eta_L \xi \eta_L^{-1}).
\label{A5}\ee
We first set the corresponding moment map to zero,
in other words introduce the first class constraints
\be
J^R_+ =0
\qquad\hbox{and}\qquad
(gJ^R g^{-1})_+ +\xi =0,
\label{A6}\ee
and then factorize by the action of $K$.
The resulting reduced phase space can be identified with the Ruijsenaars-Schneider phase space
$(M,\omega)$ (\ref{10}) by means of a global gauge slice $S$.
To describe $S$, we need to introduce an
$\cO(\chi)$-valued function on $M$ by
\be
\xi(q,p):=
\ri\chi (\1_n- v(q,p) v(q,p)^\dagger)
\quad\hbox{with}\quad
v(q, p):= L(q, p)^{-\frac{1}{2}} u(q,p),
\label{A7}\ee
where $L(q,p)$ is the Lax matrix (\ref{13}) and $u(q,p)$ is the column vector formed by the
components $u_j(q,p)$ (\ref{14}).
Denoting its elements as triples according to (\ref{A1}),
the gauge slice $S$ is  given by
\be
S:= \{ ( L(q, p)^\frac{1}{2}, -2 \mathbf{q}, \xi(q,p))\,\vert\,
(q,p)\in \cC_n\times \bR^n \}.
\label{A8}\ee
It has been shown in \cite{JPA} that $S$ is
 a global cross section of the $K$-orbits in the constraint-surface defined by (\ref{A6}),
 and the reduced Poisson brackets (alias the Dirac brackets)
 reproduce the canonical Poisson brackets (\ref{12}).

Relying on the above result, we identify $(M,\omega)$ with the model $S$ of the reduced phase space.
We can then realize the functions $I_k, I_k^1\in C^\infty(M)$ (\ref{15}) as the
restrictions to $S$ of respective $K$-invariant functions $\cI_k, \cI_k^1$ on the unreduced
phase space (\ref{A1}) furnished by
\be
\cI_k(g,J^R,\xi):= \tr \bigl((g^\dagger g)^k\bigr),
\quad
\cI_k^1(g,J^R,\xi):= -\frac{1}{2} \Re\tr \bigl((g^\dagger g)^k J^R\bigr)
\quad
\forall k\in \bZ.
\label{A9}\ee
Now the point is that the relation $\{ \cI_k^1, \cI_j\} = j \cI_{j+k}$ follows obviously from (\ref{A4}),
and this implies
(\ref{18}) by restriction to $S\simeq M$ by using that the Poisson brackets of the
$K$-invariant functions survive the reduction. The relation
\be
\{ \cI_k^1, \cI_j^1\}=(j-k) \cI_{k+j}^1
\qquad
\forall j,k\in \bZ
\ee
is also easy to confirm with the aid of (\ref{A4}), which implies (\ref{19}).

\medskip

We finish with a few remarks.
First, we note that the variables (\ref{20}) can be useful for
constructing compatible Poisson structures for the rational RS model.
In this respect,  see \cite{Falqui,Jev} and references therein.
Second, one may also construct RS versions of the Calogero
constants of motion considered
in \cite{Ran}.
Third, it could be interesting to study
quantum mechanical analogues of the constants of motion and to characterize their algebras.
It is very likely that the algebra of equations (\ref{18}), (\ref{19})
survives quantization.
One may address this question by generalizing the method
applied  in \cite{Kuz} to quantize the analogous algebra in the Calogero case.
It should be also possible to construct a quadratic algebra for the model
by suitably replacing the Dunkl operators used in \cite{Kuz} with Dunkl-Cherednik
operators. The pertinent algebras are expected to be closely
related to the bispectral property of the rational RS model \cite{Chal}.
Finally, it is natural to ask about generalizations concerning
the hyperbolic RS model and
its non-relativistic limit. We plan to return to some of these issues elsewhere.

\medskip
\bigskip
\noindent{\bf Acknowledgements.}
This work was supported
by the Hungarian
Scientific Research Fund (OTKA) under the grant K 77400.
We thank A.~Jevicki for informing us about reference \cite{Jev}, and are grateful
to our colleagues for discussions and correspondence.


\begin{thebibliography}{9}

\bibitem{CRM}
P.~Tempesta, P.~Winternitz  {\it et al} (Editors),
Superintegrability in Classical and Quantum Systems,  CRM Proceedings and Lecture Notes, {\bf 37},
 Amer. Math. Soc., 2004

\bibitem{Rag}
A.~Ballesteros, A.~Enciso, F.J.~Herranz and O.~Ragnisco,
{\it Superintegrability on N-dimensional curved spaces: Central potentials, centrifugal terms and monopoles},
Ann. Phys. {\bf 324} (2009) 1219-1233


\bibitem{Evans}
N.W.~Evans,
{\it Superintegrability in classical mechanics}, Phys. Rev. A {\bf 41} (1990) 5666-5676

\bibitem{Mill}
E.~G. Kalnins, J.~Kress and W.~Miller, Jr.,
{\it Second order superintegrable
systems in conformally flat spaces III. Three-dimensional classical
structure theory}, J. Math. Phys. {\bf 46} (2005) 103507


\bibitem{Cal}
F.~Calogero,
{\it Solution of the one-dimensional $N$-body problem with quadratic and/or
inversely quadratic pair potentials},
J. Math. Phys. {\bf 12} (1971) 419-436



\bibitem{Sut}
B.~Sutherland,
{\it Exact results for a quantum many-body problem in one dimension II},
Phys. Rev. {\bf A5} (1972) 1372-1376



\bibitem{CalRag}
F.~Calogero, O.~Ragnisco and C.~Marchioro,
{\it Exact solution of the classical and quantal one-dimensional many-body problems with
the two-body potential $V_a(x)=g^2 a^2/\sinh^2(ax)$},
Lett. Nuovo Cim. {\bf 13} (1975) 383-387

\bibitem{RS}
 S.N.M.~Ruijsenaars  and H.~Schneider,
{\it A new class of integrable models and their relation to solitons},
Ann. Phys. (N.Y.)  {\bf 170} (1986) 370-405

\bibitem{SR-CMP}
S.N.M.~Ruijsenaars,
{\it Action-angle maps and scattering theory for some finite-dimensional
integrable systems I. The pure soliton case},
Commun. Math. Phys.  {\bf 115} (1988) 127-165

\bibitem{W}
S.~Wojciechowski,
{\it Superintegrability of the Calogero-Moser system}, Phys. Lett. {\bf 95A} (1983) 279-281

\bibitem{Sas}
R.~Caseiro, J.-P.~Francoise and R.~Sasaki,
{\it Algebraic linearization of dynamics of Calogero type for any Coxeter group},
J. Math. Phys. {\bf 41} (2000) 4679-4686

\bibitem{Gon}
C.~Gonera,
{\it On the superintegrability of Calogero-Moser-Sutherland model},
J. Phys. A: Math. Gen. {\bf 31} (1998) 4465-4472

\bibitem{Tsig}
A.V.~Tsiganov,
{\it On maximally superintegrable systems},
Regul. Chaotic Dyn. {\bf 13} (2008) 178-190


\bibitem{JPA}
L.~Feh\'er and C.~Klim\v c\'\i k,
{\it On the duality between the hyperbolic Sutherland and the rational Ruijsenaars-Schneider models},
J. Phys. A: Math. Theor. {\bf 42} (2009)  185202


\bibitem{Falqui}
C.~Bartocci, G.~Falqui, I.~Mencattini, G.~Ortenzi and M.~Pedroni,
{\it On the geometric origin of the bi-Hamiltonian structure of the Calogero-Moser system},
arXiv:0902.0953 [math-ph]

\bibitem{Jev}
I.~Aniceto, J.~Avan and A.~Jevicki,
{\it Poisson structures of Calogero-Moser and  Ruijsenaars-Schneider models},
arXiv:0912.3468 [hep-th]

\bibitem{Ran}
M.F.~Ranada,
{\it Superintegrability of the  Calogero-Moser system:
Constants of motion, master symmetries, and time-dependent symmetries},
J. Math. Phys. {\bf 40} (1999) 236-247

\bibitem{Kuz}
V.B.~Kuznetsov,
{\it Hidden symmetry of the quantum Calogero-Moser system},
Phys. Lett. A {\bf 218} (1996) 212-222

\bibitem{Chal}
O.A.~Chalykh,
{\it Bispectrality for the quantum Ruijsenaars model
and its integrable deformation},
Journ. Math. Phys. {\bf 41} (2000) 5139-5167


\end{thebibliography}
\end{document}